\documentclass[12pt,preprint]{aastex}

\shorttitle{Spizer Observations of two TWA BDs}
\shortauthors{Riaz, Gizis & Hmiel}

\begin{document}

\title{Spitzer Observations of two TW Hydrae Association Brown Dwarfs}

\author{Basmah Riaz\altaffilmark{1}, John E. Gizis and Abraham Hmiel}
\affil{Department of Physics and Astronomy, University of Delaware,
    Newark, DE 19716; basmah@udel.edu, gizis@udel.edu, abehmiel@udel.edu}

\altaffiltext{1}{Visiting Graduate Research Fellow, Spitzer Science Center, California Institute of Technology, Pasadena, CA 91125; basmah@ipac.caltech.edu}

\begin{abstract}
We present Spitzer Space Telescope observations of two TW Hydrae Association brown dwarfs, 2MASSW J1207334-393254 and 2MASSW J1139511-315921, in the IRAC and MIPS 24 $\micron$ bands. Based on their IRAC colors, we have classified them as Classical and Weak line T Tauri stars, respectively. For 2MASSW J1207334-393254, we have found that a flat disk model fits the data very well. This brown dwarf shows the presence of warm (T $\ga$ 100 K) circumstellar dust close (R $\la$ 0.2 AU) to it, and does not display any signs of cleansing of dust within several AU of the star. In comparison with other TWA members that show excess in IR, we suggest that there exists a different disk evolution/dust processing mechanism for stellar and sub-stellar objects. 2MASSW J1139511-315921 does not show any significant excess in any of the IRAC bands, but a small one at 24 $\micron$, which is not significant enough to suggest the presence of warm dust around this star. It shows signs of dust cleansing in the inner several AU, similar to most of the other TWA members.
\end{abstract}


\section{Introduction}

TW Hydrae Association (TWA) has been under observation since its identification by Kastner et. al. (1997). This young ($\thicksim$10Myr, Webb et. al. 1999) and nearby ($\thicksim$50pc) association now has more than 25 confirmed members, in both stellar and sub-stellar regimes, with most displaying signatures of being in the T Tauri phase. Consisting mainly of stars that vary from late K down to the brown dwarf masses, TWA has proven to be an important field to study formation and evolutionary differences, if any, in stellar and sub-stellar objects, and their activity. A tool for such a study has been the detection of excess in the IR and Far-IR bands. This has helped in classifying them as T Tauri stars with disks or young stars without disks. An absence of such excess suggests the cleansing of warm dust around the star due to the presence of a planetary companion (Weinberger et. al. 2004) or strong stellar winds. Low et. al. (2005) (hereafter, L05), observed 24 TWA members using Spitzer and found measurable IR excess in only 6 stars, concluding on the absence of warm dust around these young stars. The two strong accretors, TWA Hya and Hen 3-600 show strong IR excess, as expected. TWA 7 and TWA 13 show excess only at or after 70 $\micron$ (L05). Hartmann et. al. (2005) (hereafter, H05) conducted IRAC observations of brown dwarfs in Taurus, and found a strong correlation between excess in the IRAC bands and the presence of a disk around the star, which helped in classifying them as Weak line T Tauri Stars (WTTS) or Classical T Tauri Stars (CTTS). Sterzik et. al. (2004) (hereafter, S04), reported excess around both a $\thicksim$1Myr old brown dwarf Cha H$\alpha$ 1 and the $\thicksim$10Myr old 2MASSW J1207334-393254. Luhman et. al. (2005a) discovered a disk around the least massive known brown dwarf OTS 44. The variations in the SEDs of TWA stars, as seen by L05, suggests that the end of the T Tauri phase, and a transition from this phase to a quiescent one may not be similar for all of them.

This work is based on Spitzer Space Telescope observations in the IRAC and MIPS bands of two TWA brown dwarfs-2MASSW J1207334-393254 (hereafter, 2M1207) and 2MASSW J1139511-315921 (hereafter, 2M1139), both discovered by Gizis (2002) (hereafter, G02). These were confirmed as TWA members recently by Scholz et. al. (2005a), based on accurate proper motions. G02 classified them as M8, with masses $\thicksim 0.025 M_{\sun}$. Both show low surface gravity signatures, which confirms their youth (G02). The low surface gravity of 2M1139 was later confirmed by Gorlova et. al. (2003) based on the weakness of FeH and K I lines in the J-band and of Na I line in the K-band spectra. Mohanty et. al. (2003) confirmed the same for both 2M1207 and 2M1139 based on narrow Na I absorption profiles. They also detected Li I (6708 $\AA$) in both stars. G02 reported very strong H$\alpha$ emission in 2M1207 (eq. width = 300 $\AA$) but a weaker one in 2M1139 (eq. width = 9.7 $\AA$). Later, Mohanty et. al. (2003) found an H$\alpha$ eq. width of 27.7 $\AA$ for 2M1207 and 7.3 $\AA$ for 2M1139. They classified 2M1139 as a non-accretor, chromospherically active WTTS-like dwarf. 2M1139 failed to be detected in the radio by Burgasser \& Putman (2005). 2M1207, on the other hand, was classified as an CTTS-like accretor, based on the detection of He I and upper Balmer lines and the condition that accretors should display broad asymmetric H$\alpha$ emission (Mohanty et. al. 2003). This variation in 2M1207's H$\alpha$ emission and its accretory nature was confirmed by Scholz et. al. (2005b), who found that the accretion rate varies by a factor of 5-10 on a timescale of 6 weeks. Gizis $\&$ Bharat (2004) found an absence of X-ray emission in 2M1207, and concluded that the chromosphere and the transition region are weak in this brown dwarf. Chauvin et. al. (2004; 2005) discovered a planetary mass companion candidate that shares the same proper motion as 2M1207, and lies at a separation of $\thicksim$55AU. Its spectral type, mass and effective temperature were determined by these authors to be L5-L9.5, $M = 5 \pm 2 M_{JUP}$, $T_{eff} = 1250 \pm 200$ K, respectively. Gizis et. al. (2005) studied 2M1207 in the UV and concluded that it is accreting actively from a circumstellar gas disk by identifying CIV line as due to accretion and H$_2$ lines due to the circumstellar gas.

To check any excess in the IR due to the presence of a disk, Jayawardhana et. al. (2003) observed 2M1207 and 2M1139 in the L$^{\prime}$-band, but failed to detect any K-L$^{\prime}$ excess. Based on the redshifted absorption component in the H$\alpha$ profile, Scholz et. al. (2005b) concluded that 2M1207 is seen with an inclination of $i \geqslant  60\degr$. They thus related the lack of K-L$^{\prime}$ excess in 2M1207 to a high inclination disk. S04 found excess at 8.7 $\micron$ and 10.4 $\micron$ and detected the presence of a dusty disk around 2M1207. However, due to the unavailability of ISOCAM data and lack of excess in the L$^{\prime}$-band, there were just two data points available to determine the shape and strength of the disk. S04 concluded that both a large grain size, large inclination angle flared disk, and a small grain size flat disk fit the two data points well. We have observed 2M1207 in the Spitzer (Werner et al. 2004) IRAC (Fazio et al. 2004) and MIPS (Rieke et al. 2004) 24 $\micron$ bands in order to characterize the disk.

\section{Observations}

The observations were made for the two stars in all four IRAC bands and MIPS 24 $\micron$ band only. IRAC has a plate scale of 1.22$\arcsec$ $\slash$ pixel and a field of view of 5.2$\arcmin$ $\times$ 5.2$\arcmin$. We used the 12s high dynamic range mode with 5 dithers in each band to improve the S/N. The IRAC observations were made on the 10th and 14th of June, 2005. The MIPS 24 $\micron$ photometry was obtained for both stars on the 27th of January and 1st of February 2005. The MIPS 24 $\micron$ plate scale is 2.45$\arcsec$ $\slash$ pixel. We used an exposure time of 10 s and 6 template cycles to ensure high S/N.

Aperture photometry was performed on the artifact-corrected mosaic images using the task PHOT under the IRAF package APPHOT. We used an aperture radius of 3 pixels and a background annulus of 3-7 pixels for the IRAC data. The zero point fluxes of 280.9, 179.7, 115 and 64.1 Jy and aperture corrections of 1.124, 1.127, 1.143, and 1.234 were used for IRAC channels 1 through 4, respectively. The errors in magnitudes for 2M1207 are $\thicksim$0.08 mag for IRAC channels 1 and 2, and $\thicksim$0.1 mag for channels 3 and 4. For 2M1139, the photometric errors are $\thicksim$0.07 mag for IRAC channels 1 and 2, and $\thicksim$0.11 mag for channels 3 and 4. For the MIPS 24 $\micron$ data, an aperture radius of 6 pixels, background annulus of 12-17 pixels, zero point flux of 7.3 Jy and an aperture correction of 1.19 were used. The magnitude errors are $\thicksim$0.1 mag for 2M1207 and $\thicksim$0.22 mag for 2M1139. The fluxes thus determined are listed in Table 1.

\section{Discussion}

\subsection{Spectral Energy Distributions (SEDs)}

Figures 1 and 2 show the SEDs of 2M1207 and 2M1139, respectively. We have used the BD Dusty98 model (Leggett et. al. 1998) for 2M1207 and the NextGen model (Hauschildt et. al. 1999a,b) for 2M1139. The difference between the two is that the BD Dusty models take into account the dust opacities that can affect the equivalent widths of the spectral features of the very cool objects. But significant differences become apparent only below $T_{eff}  \thicksim 2300$K (Gorlova et. al. 2003). We have found a good fit for 2M1207 and 2M1139 at a $T_{eff}$ of 2400K and 2600K, respectively. The radii are similar, $R_{1207} = 0.33 R_{\sun}$ and $R_{1139} = 0.35 R_{\sun}$. We have used a distance of 70pc for 2M1207 (S04) and 73pc for 2M1139 (Gorlova et. al. 2003).   Gorlova et. al. (2003), calculated a value of $\log g = 3.62$ for 2M1139. We have used the closest value the models could offer ($\log g = 3.5$) for both stars.   The temperature and radius fits are illustrative: adequate for our purposes of detecting disk excesses but not meant to be definitive.  

As can be seen from Fig. 1, 2M1207 shows significant excess in all of the IRAC bands and at 24 $\micron$. Fig. 1 shows that the flat disk model fits well to most of the data points that show excess in emission compared to the photosphere. We have used a straight line of slope $\thicksim$ -1.3 to fit these data points. A steep spectral slope as this one suggests that the disk becomes optically thin at longer wavelengths. Flat disk geometry is independent of the grain size, thus a small or large grain size wouldn't matter, as can also be seen from Figure 4 in S04. Also, for $\lambda \la 9 \micron$, the maximum grain size does not matter (Luhman et. al. 2005a).  The excess at 10 microns is due to silicate emission, as argued by S04; Gizis et al. (2005) found that the accreting gas is depleted in silicon.  

Jayawardhana et. al. (2003) reported an absence of excess for 2M1207 at 3.8 $\micron$, and suggested that the reason could be an inner hole in the disk. It should be noted that they used a conservative limit of K-L$^ {\prime} \ga 0.2$ to define this excess. Scholz et. al. (2005b) related this lack of K-L$^{\prime}$ excess due to a high inclination disk ($i \geqslant 60 \degr$). However, we have a flat disk model fit now, which shouldn't be affected by the inclination angle.  We have not detected a considerable excess at 3.6 $\micron$, and therefore support the idea of an inner hole.  

Recently, Scholz et. al. (2005c) have shown that accreting sub-stellar objects may have magnetic fields of $\thicksim$kG strengths. Magnetic fields of such strengths can perturb the accretion disk, causing accreting material to fall at free-fall velocities onto the star. If the inner disk radius is co-rotating with the magnetic field, the dissipation of energy as the magnetic field interacts with the disk is not going to be very high. The inner hole deduced by the lack of a significant emission at 3.6 $\micron$ corresponds to the co-rotation radius of the disk, if the temperature of the disk is dominated by irradiation. This is where we expect the magnetic fields to disrupt the disk.

L05 suggested that the excess at 24 $\micron$ places severe limits on the presence of warm (T $\ga$ 100K) circumstellar dust around the star. The observed flux for 2M1207 at 24 $\micron$ is 11 times more than what the photosphere predicts. This suggests the presence of warm dust around the star. Using the lower limit on $T \ga 100$K, we find an upper limit on the minimum distance from the star for dust to be in thermal equilibrium to be $R \la 0.2$ AU. We note that based on the extent of excess at 24 $\micron$, the upper limit on R could be smaller than the one calculated above. Thus 2M1207 shows strong signs of warm dust close to it.

Fig. 2 shows the SED of 2M1139. We have not detected any significant excess at any wavelength except a small one at 24 $\micron$. It has been classified as a non-accretor, chromospherically active star by Mohanty et. al. (2003). It does not show much H$\alpha$ variability (H$\alpha$ eq. width = 9.7 $\AA$ (G02) and 7.3 $\AA$ (Mohanty et. al. 2003)). Jayawardhana et. al. (2003) couldn't detect any excess at 3.8 $\micron$. Based on its IRAC colors (see discussion under color-color plots), and the absence of a disk, we conclude that it is a WTTS-analog brown dwarf. Since the excess at 24 $\micron$ is very small, it is unlikely that there is warm dust present around the star. 2M1139 shows cleansing of dust within several AU of the star, similar to other TWA members. However, what needs to be checked is whether 2M1139 will display any excess at 70 $\micron$ and longer wavelengths or not. As L05 observed, a star may display an excess at 70 $\micron$, even if it shows none at 24 $\micron$. 2M1139 might turn out to be a case similar to TWA 7, for which cooler ($\thicksim$80 K) material has been detected at $R \geqslant 7$ AU from the star (L05). 

Weinberger et. al. (2004) suggested that rapid planet formation within 5-10 Myr is the reason why the K and M dwarfs in the TWA show a lack of warm circumstellar dust around them, displayed by an absence of infrared excess. Out of the 24 stars in their sample, L05 found excess in only 6 of them, 4 of which are confirmed binary systems. The absence of excess within 24 $\micron$ in 2M1139 and TWA 7 suggests the presence of an unseen companion. Recently, Scholz et. al. (2005a) have discovered a $\thicksim 20 M_{JUP}$ mass, probable new TWA member, SSSPM J1102-3431, that lies only 12$\arcmin$ from TW Hya, and conclude that the two possibly form a wide binary system with a separation of 40,000 AU at $d = 56$pc. TW Hya shows a small excess at 8 $\micron$ and none short ward of that wavelength (H05). If we try to estimate the excess in the IRAC bands for L05's 6 stars, none would display any excess short ward of 8 $\micron$. The fact that 2M1207 has a distant planetary mass companion, and yet displays the presence of warm dust close to it, is intriguing. This could be interpreted that 2M1207 is not coeval with the other TWA members, but that seems unlikely since a large variety is seen among the SEDs of TWA stars (L05). Also, even among brown dwarfs, disk evolution is not uniquely a function of age. At the same age of $\thicksim$1 Myr, Cha H$\alpha$ 1 shows the presence of a flared small grain size disk (S04), whereas an optically flat disk fits the data well for Cha H$\alpha$ 2 (Apai et. al. 2002). Apai et. al. (2002) concluded that BDs display unexpected differences in their disk geometry. Thus at the same age, 2M1207 could display lesser dust processing than the other TWA members. Another explanation for the simultaneous presence of warm dust and a planet around 2M1207 could be that the six TWA stars in L05's sample with confirmed IR excess are all late K- to early M- type stars. This could suggest a different dust processing/disk evolution mechanism for early and late type stars. Indeed, Plavchan et. al. (2005) have concluded that there exists a different debris disk evolution mechanism for K and M dwarfs, based on the absence of an excess at 11.7 $\micron$ in their sample of nine M dwarfs ($T_{eff}  \ga 3000$K), and that the M dwarf debris disks dissipate within 5-10 Myr. L05 suggested that dust is swept away within several AU of the star in its early days of evolution. At an age of $\thicksim$10 Myr, 2M1207 shows weak signs of dissipation of the dust close to the star; however, the coeval 2M1139 has dissipated its dust. L05 concluded that in TWA K- and M- type stars, the transition period from the T Tauri phase to a quiescent phase is very short. 2M1207 still displays T Tauri like activity. These arguments suggest that although the formation of stellar and sub-stellar objects may be similar, the disk evolution/dust processing mechanisms may be different. The onset of the T Tauri phase may occur at the same time, but the timescale for the transition from this phase to a quiescent one may not be the same. We also note that if the planet around 2M1207 did not form in the disk, as suggested by Chauvin et. al. (2005), and given its great distance, then it would be expected to have little effect on the disk.  

We did not request for observations of 2M1207 and 2M1139 in the MIPS 70 and 160 $\micron$ bands.  Extrapolating from 24 $\micron$ onwards, we predict excesses for 2M1207 at 70 and 160 $\micron$ too. We have calculated the disk temperature, $T_{d}$, for 2M1207 as a function of the inner radius, R, using the equation, $T_{d}^{4}(R) \thicksim (T_{eff}^4 / \pi)(R_{star} / R)^{3}$. The results show that at around 70 and 160 $\micron$, the disk temperature will be less than $\thicksim$40K, suggesting that the flux will be less than $\thicksim$10 times of what the photosphere predicts (in comparison with TWA 13, L05). At $\thicksim$45 AU, the temperature comes out to be $\thicksim$0.8K. Thus the outer cutoff due to the planetary companion ($\thicksim$55AU) around 2M1207 does not have any effect on our Spitzer observations since the temperatures are too low out there. In any case, observations in the MIPS 70 and 160 $\micron$ bands will be valuable in evaluating the outer radius of the disk.

\subsection{Color-Color Plot}

Fig. 3 shows the color-color plot for the IRAC bands. We have added some early type field M dwarfs (results from a work in progress) which are all WTTS, and IRAC observations of Cha H$\alpha$ 1 (Luhman et. al. (2005b)) and TW Hya (H05). H05 have used a limit of [5.8]-[8] $\thicksim$ 0.3-0.4 as a limit to differentiate between CTTS and WTTS. This criteria defines 2M1207 as a CTTS and 2M1139 as a WTTS, respectively. Another limit at [3.6]-[4.5] $\thicksim$ 0.2 could be used to separate the CTTS with small or large inner holes in their disks. 2M1207 and Cha H$\alpha$ 1 show signs of a small inner hole and lie above this limit. Whereas, TW Hya, that has a larger inner hole, lies below this limit. We note that these limits are based on just three stars. However, by looking at the SEDs of L05's stars (like TWA 3, TWA 4 and TWA 11), we can be sure that these will display very small or no excess in the IRAC bands, which would certify these limits.

\section{Summary}

Based on their IRAC colors, we have classified 2M1207 as a CTTS-like brown dwarf and 2M1139 as a WTTS-like brown dwarf, as already suggested by their H$\alpha$ line profiles. We have found that a flat disk model fits the excess well for 2M1207.  This brown dwarf shows little sign of cleansing of dust close to the star, based on the excess shown in the IRAC and MIPS 24 $\micron$ bands. The excess at 24 $\micron$ certifies the presence of warm (T $\ga$ 100K) circumstellar dust close (R $\la$ 0.2AU) to the star. In comparison with other TWA members that show excess in IR, we suggest that there exists a different disk evolution/dust processing mechanism for stellar and sub-stellar objects. We predict, based on the excess at 24 $\micron$, that 2M1207 will show excess at 70 and 160 $\micron$ too. 2M1139 shows insignificant excess in all of the IRAC bands, and a small one at MIPS 24 $\micron$. It shows signs of dust cleansing in the inner several AU, similar to most of the other TWA members. The absence of warm dust around it suggests the possible presence of a planetary companion.

\acknowledgments

This work is based on observations made with the Spitzer Space Telescope, which is operated by the Jet Propulsion Laboratory, California Institute of Technology under a contract with NASA. Support for this work was provided by NASA through an award issued by JPL/Caltech.

{\it Facilities:} \facility{Spitzer Space Telescope}

\clearpage

\begin{figure}
\resizebox{150mm}{!}{\includegraphics[angle=270]{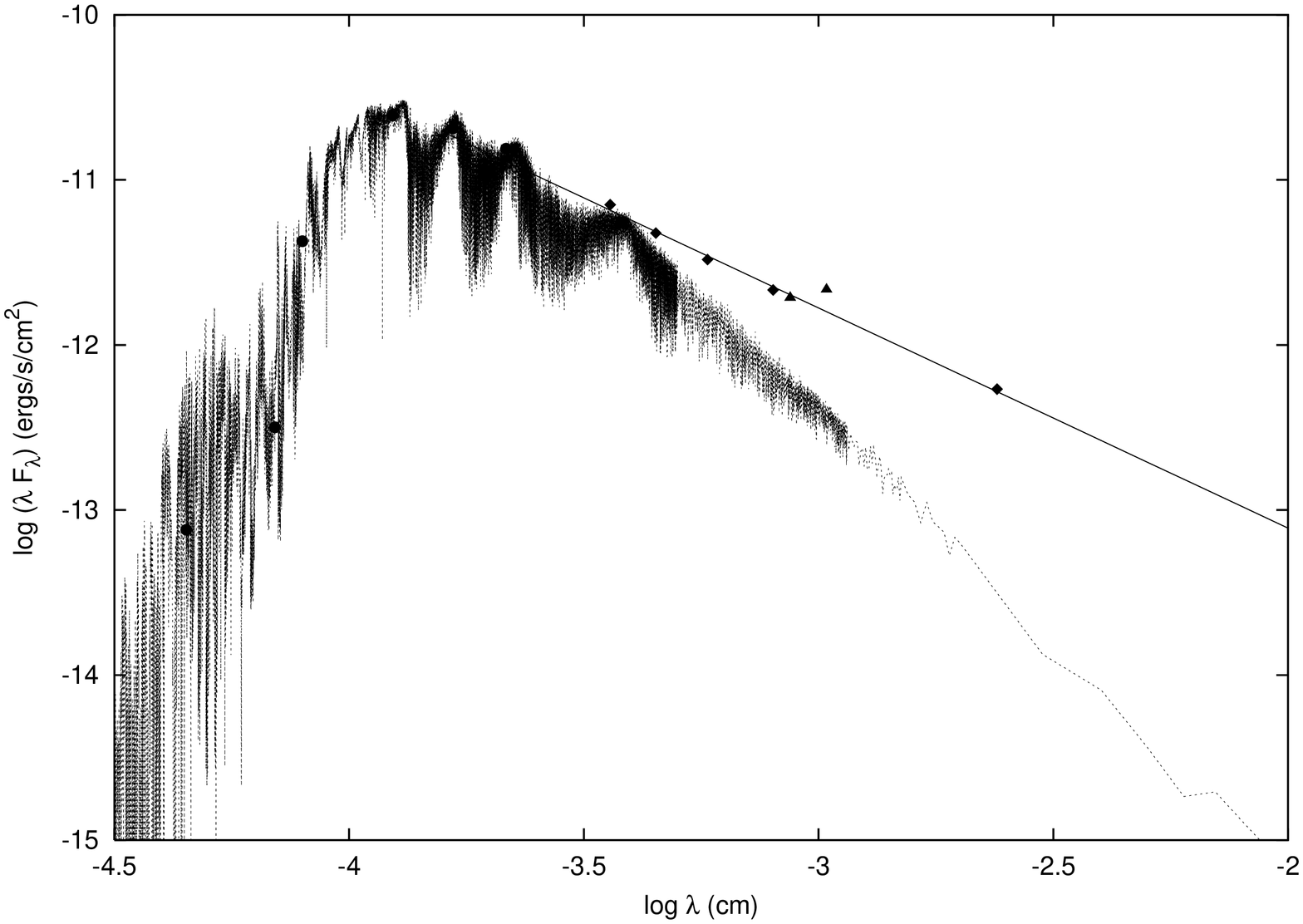}}
\caption{2M1207 SED. Filled circles represent optical and J, H, K magnitudes obtained from DENIS and 2MASS, respectively. Filled diamonds (this work), filled square (Jayawardhana et. al. 2003) and filled triangles (S04). Dotted line represents the BD Dusty98 model for a $T_{eff} = 2400$K, $R = 0.33 R_{\sun}$ and $\log g=3.5$. Solid line represents a straight line fit of slope -1.3 to the data points that show excess in emission.}
\end{figure}

\clearpage

\begin{figure}
\resizebox{150mm}{!}{\includegraphics[angle=270]{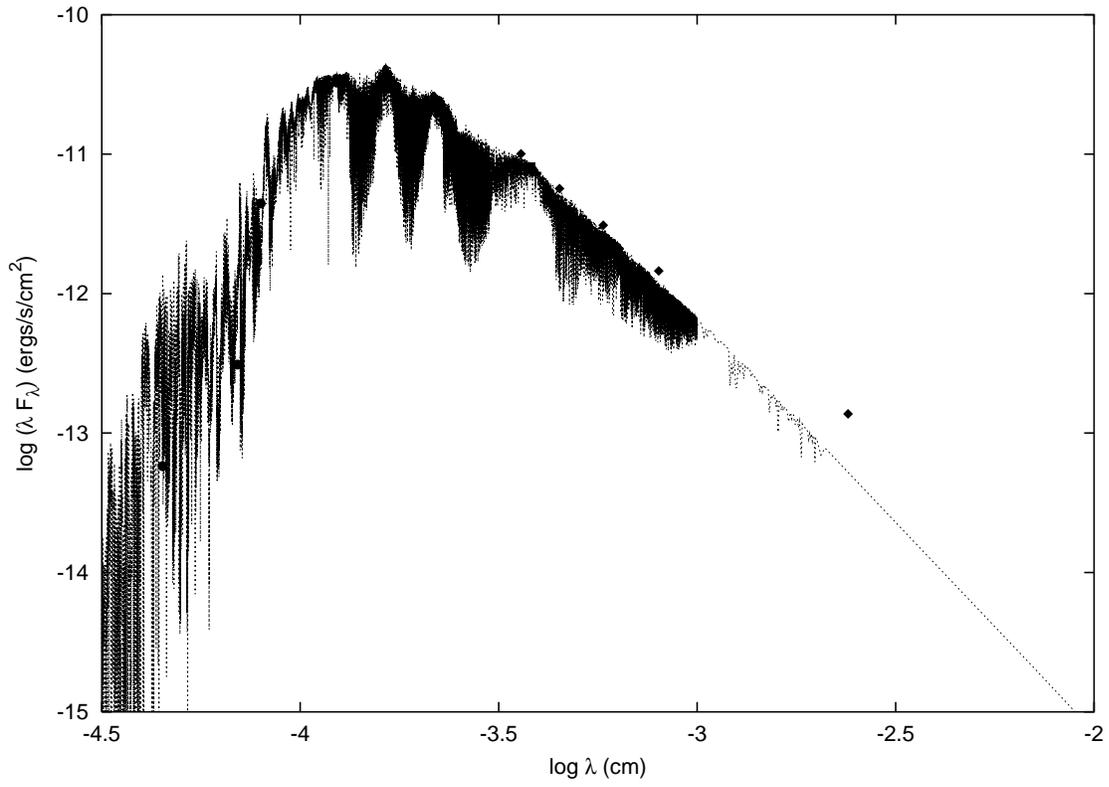}}
\caption{2M1139 SED. Symbols are the same as in Fig. 1, except dotted line represents the NextGen model, for a $T_{eff} = 2600$K, $R = 0.35 R_{\sun}$ and $\log g=3.5$}
\end{figure}

\clearpage

\begin{figure}
\resizebox{150mm}{!}{\includegraphics[angle=270]{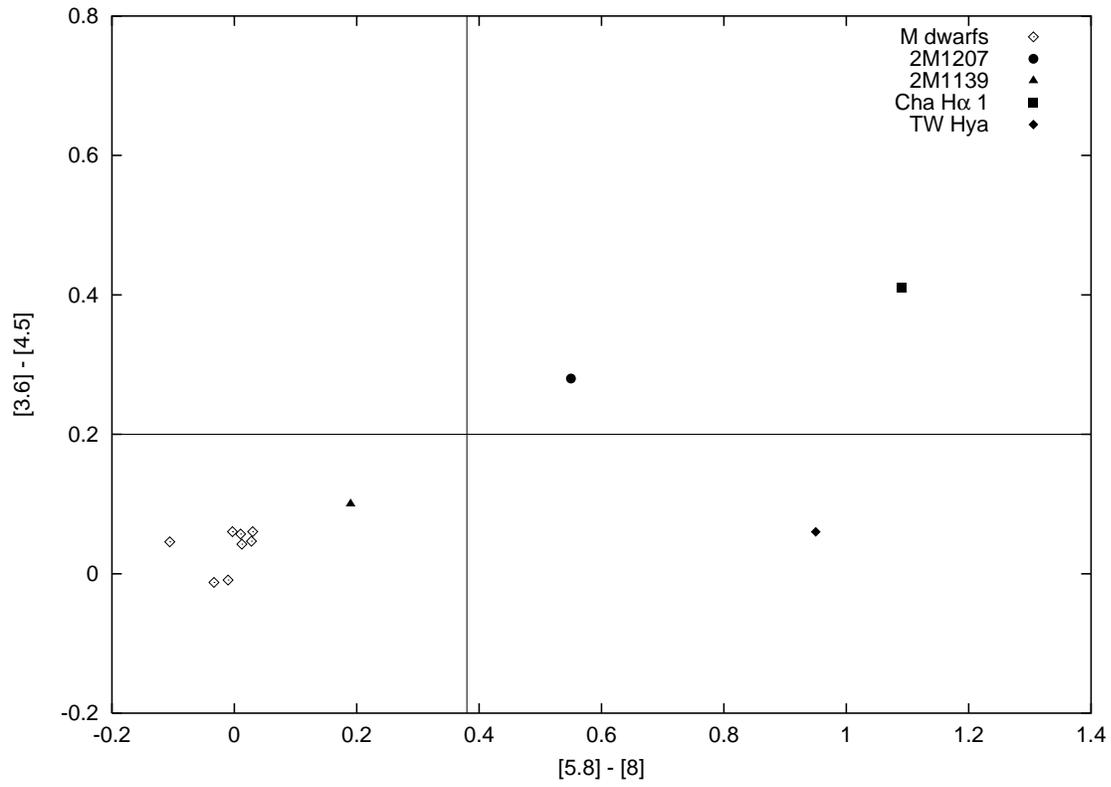}}
\caption{IRAC color-color plot. The field M dwarfs have spectral types between M2-M5. The spectral types of Cha H$\alpha$ 1 and TW Hya are M7.5 and K8, respectively.}
\end{figure}

\clearpage

\clearpage

\begin{deluxetable}{ccrrrrrrrrcrrl}
\tabletypesize{\scriptsize}
\tablecaption{IRAC and MIPS Observations}
\tablewidth{0pt}
\tablehead{
\colhead{Star} & \colhead{3.6 $\micron$} & \colhead{S/N} & \colhead{4.5 $\micron$} & \colhead{S/N} &
\colhead{5.8 $\micron$} & \colhead{S/N} & \colhead{8 $\micron$} &
\colhead{S/N} & \colhead{24 $\micron$} &
\colhead{S/N} & \colhead{3.8 $\micron$\tablenotemark{a}} & \colhead{8.7 $\micron$\tablenotemark{b}} & 
\colhead{10.4 $\micron$\tablenotemark{b}} \\
\colhead{} & \colhead{mJy} & \colhead{} & \colhead{mJy} & \colhead{} & \colhead{mJy} &
\colhead{} & \colhead{mJy} & \colhead{} & \colhead{mJy} & \colhead{} & \colhead{mJy} &
\colhead{mJy} & \colhead{mJy}  
}
\startdata
2M1207 & 8.49 & 405 & 7.15 & 372 &  6.36 & 350 & 5.74 & 332 & 4.32 & $>$1000 & 7.00 & 5.60 & 7.50  \\
2M1139 & 12.08 & 509 & 8.51 & 427 & 5.98 & 358 & 3.88 & 288 & 1.10 & 860 & 10.23 & - & - \\
\enddata

\tablenotetext{a}{Jayawardhana et. al. (2003)}
\tablenotetext{b}{Sterzik et. al. (2004)}

\end{deluxetable}

\end{document}